\documentstyle[aps,prl,multicol,epsf,floats,epsfig,amsmath,amssymb]{revtex}

\begin{document}
  
 \draft 

 \title
{Discrete Spectrum in Bose Condensation}
\author {Enrico Celeghini $^1$ and Mario Rasetti $^2$}
\address{$^1$ Dipartimento di Fisica, and Sezione INFN, \\
Universit\`a di Firenze, I-50125 Firenze, Italy \\
$^2$ Dipartimento di Fisica, and Unit\`a INFM, \\
Politecnico di Torino, I-10129 Torino, Italy }

 \maketitle

 \begin{abstract}
{A quantitative description of the whole process of condensation of bosons 
in an harmonic trap is given resorting only to Gibbs and Bose postulates, without 
assuming equipartition nor continuum statistics.  Below $T_c$ discrete spectrum 
theory predicts for the thermo-dynamical variables a behavior different from the 
continuum case. In particular a new critical temperature $T_d$ is found   
where the specific heat exhibits a $\lambda$-like spike.  Numerical values of the 
relevant quantities depend on the experimental set-up. The theory does not take 
into account collective effects in the ground state.}
\end{abstract}


\begin{multicols}{2}[]

\narrowtext

Bose-Einstein condensation (BEC) was discovered (or, better, predicted) by 
Bose \cite{Bos} and Einstein \cite{Ein} in 1924-25 studying  the properties of 
a gas of non interacting bosons in a (infinite) box.  BEC has been experimentally 
observed only in 1995, in vapors of rubidium \cite{And} and sodium \cite{Dav} 
confined in magnetic traps and cooled. Present research on BEC follows similar 
experimental lines, where the potential makes the spectrum discrete \cite{GSS}, 
\cite{Inguscio}.  Theoretical research on BEC in traps is also in tumultuous 
progress \cite{DGPS}, focusing essentially on the role of interactions among the 
atoms. Here we shall instead pay attention more to the relation between the discreteness 
and multiplicity of the quantum mechanical spectrum, and the features induced by 
quantum statistics.  

BEC in a box (where the walls do not essentially affect the 
continuity of the spectrum) and BEC in an harmonic trap (whose effect is to make the 
spectrum definitely discrete), coincide only in some approximate way. 
First of all, in a magnetic trap the number $N$ of confined particles as well as the 
volume $V$ in which they are confined are finite, namely we are not in the thermodynamic 
limit. This effect has been discussed in different papers \cite{KD}, \cite{GH}, \cite{KT}, 
\cite{HHR} and shown to have limited phenomenological relevance, as in the experiments 
the number of particles is $10^5\div 10^7$.  There is however another reason why the 
thermodynamic limit is not reached -- and cannot be reached -- in magnetic traps, on 
which we focus our attention here: the energy spectrum. Quantizing in a box, the difference 
between contiguous energy levels goes to zero at least as $V^{-1/3}$ as volume increases. 
On the other hand, in order to obtain the thermodynamic limit in an harmonic potential we 
should have to require that the oscillator frequency -- {\sl i.e.} the spacing of the 
spectrum -- goes to zero as $N^{-1/3}$.  One needs, in other words, that the spectrum 
may be considered continuous.  This is far from the experimental situation, where 
typically $h\nu /k_B \approx 10\, nK$ (for $\nu \approx 10^2 \, Hz$), to be compared with 
the estimated temperature $T_c \approx \, 10^2 nK$. 

We shall attempt here to propose a better framework within which experimental results are 
consistently described, trying to represent more properly BEC experimental situation ($N$ 
finite, energy discrete).  Even though keeping spectral structure into account may be an 
important step forward with respect to the continuum theory, many body effects are still not 
considered. The results are thus correct up to the detailed structure {\sl e.g.} of the ground 
state, as it could be described by a collective perturbation in the space of free bosons, as 
in Gross-Pitaevsky approach \cite{GSS}, \cite{DGPS}.   

Ever since 1907 Einstein himself \cite{Ein907} had pointed out how classical results on 
equipartition of energy are no longer valid for $T < h \nu /k_B$ for the quantized harmonic 
oscillator in the frame of Boltzmann statistics. This has been so far disregarded, in spite 
also of the change in statistics, as $T < h \nu /k_B$ is below the range of experiments. Several 
other authors have on the other hand discussed BEC in the context of the present experimental 
situation using equipartition together with the continuum statistics (eq. 
(\ref{cont}) below) \cite{Ensher}, \cite{BPK}, \cite{dHT}, finding that the temperature scale 
of the process is $\propto N^{1/3}$.  Within the approach presented here this scaling law, essential 
for the coherence of the scheme, is recovered, among other novel results. The 
crucial difference is however that it is obtained by re-deriving the entire theory of bosons 
with discrete spectrum, based only on the Gibbs and Bose postulates.  The central results are  
that a new equation (eq. (\ref{inv}) below) replaces the continuum distribution function, equipartition 
is valid only for $T > T_c$, and a new critical temperature is found.

Our motivation for this construction is the recent definition of bosons as coherent states of 
the algebra $su(1,1)$, given in \cite{CR}. In that paper a possible generalization of the Bose 
definition of bosons has been considered. Here this is used only to define the undegenerate states 
of the harmonic oscillator as the limit for $g \to 1$ of the general formula ( eq. (\ref{psi}) below). 

We begin by summarizing the most relevant results that a thorough numerical analysis performed 
within this conceptual scheme leads to, leaving for the second part of the paper some clarifying 
details. There turn out to be seven essential features when comparing the continuum and discrete 
cases: 

1) The formula (eq. (\ref{inv}) below) for the occupation number $n(\epsilon_i )$ as a function of 
the single particle discrete energy $\epsilon_i$ is the same as in the thermodynamic limit at zero energy,  
moreover for discrete spectrum it does not decrease exponentially as $\epsilon \to \infty$ but rather it 
becomes identically zero for $\epsilon > \epsilon_{max}$ (where $\epsilon_{max}$ has a well defined 
value much less than the total energy $E$ and goes to infinity only in the continuum limit). 
Measurable effects are related with the product of $n(\epsilon_i )$ with the multiplicity of 
energy levels $m (\epsilon_i )$.  Fig. 1 displays the product $n(\epsilon_i )\, m(\epsilon_i )$ {\sl vs.} 
$i$. 
\hspace{1cm}

\begin{center}
\epsfig{file=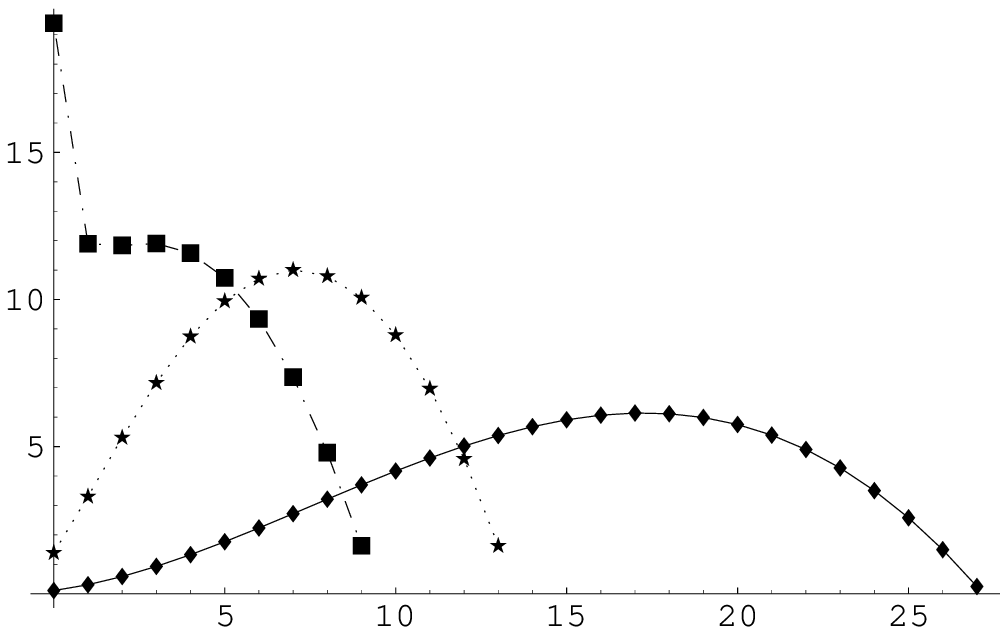,height=5cm}
\end{center}

{\noindent \footnotesize{FIG. 1. $n(\epsilon_i )\, m(\epsilon_i )$ {\sl vs.} $i$ for $N=10^2$ 
bosons in an isotropic harmonic trap for different values 
of $T$ ($T_{\blacksquare} < T_{\bigstar} < T_{\blacklozenge}$) in proximity of $T_c$, showing 
the appearance of $n_0 \neq 0$. }}

\hspace{1cm}

2) The Lagrange multiplier $\beta'$, that in the continuum is identically equal to 
$1/(k_B T)$ -- and $\propto (E/N)^{-1}$ due to equipartition -- is now for $T > T_c$ $\beta' \propto 
(E/N)^{-4}$, yet the equipartition theorem is still satisfied (constant specific heat). In other words, 
the Lagrange multiplier turns out to be a more complicated function of $T$ than $\propto T^{-1}$.  

3) Temperature, defined -- without resorting to equipartition -- as the derivative of energy 
with respect to entropy, allows us to describe entirely the thermodynamics of the system,  
assumed in global thermal equilibrium (as done in \cite{Ensher}) because of the efficiency of 
evaporative cooling.  For $T \ge T_c$, effects related to the spectrum discreteness appear to 
be irrelevant, as they are in determining $T_c$ as a function of $N$ \cite{BPK}, \cite{dHT}. 
Another critical temperature $T_d$ is however found (see Fig. 2) defined by the 
ground state filling $n_0/N$ crossing the value $1/2$. 

As Fig. 2 shows, $C$ is constant, equal to $3 k_B$ (as it should, since the system does 
not condense and behaves as a classical gas) for $T > T_c$; it is almost zero (as entropy 
and energy decrease with quite different slopes because only a few atoms progressively migrate 
from the warmer tail of the spectrum into the fundamental level) for $T_d < T < 
T_c$, and exhibits a peak for $T = T_d$, where condensation becomes a global effect. 
Note that, after the spike at $T_d$, the specific heat goes to zero as $T$ goes to zero, 
as required by Nernst's theorem.  
\hspace{1cm}

\begin{center}
\epsfig{file=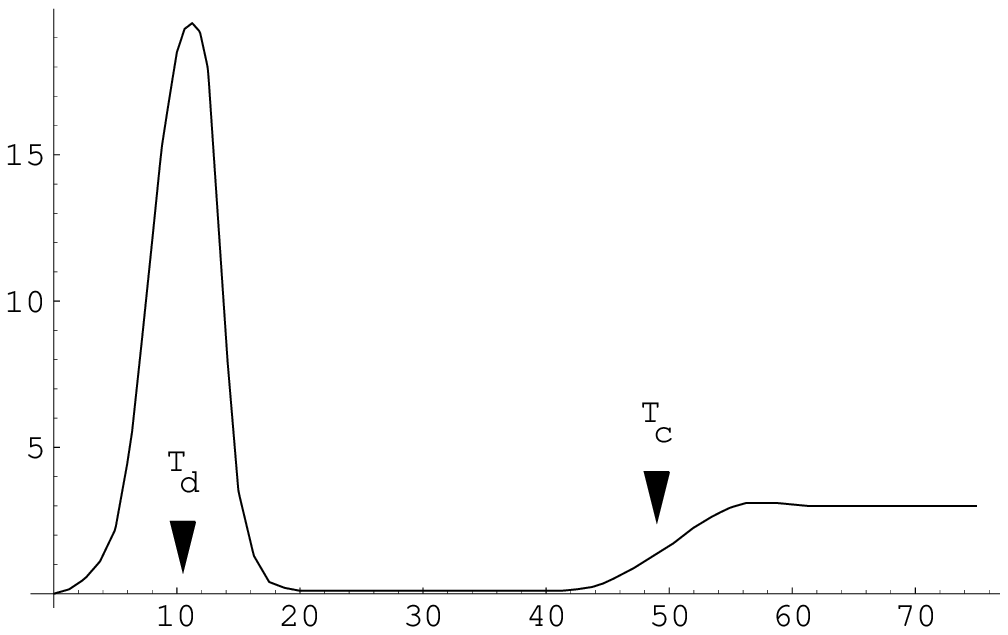,height=5cm}
\end{center}

{\noindent \footnotesize{FIG. 2. Specific heat $C$ {\sl vs.} $T$ in units $k_B = h \nu =1$ for 
$N=10^6$ bosons in an isotropic harmonic trap, as predicted by discrete quantum statistics. }}

\hspace{1cm}
  
4) The chemical potential is no longer equal to the Lagrange multiplier 
$\alpha'$ and does not appear to present major differences 
from the results reported in the literature \cite{KD}. 

5) Repeated numerical analysis for various values of $N$ indicates that $T_c \propto N^{1/3}$, in 
agreement with \cite{BPK}, \cite{dHT}, whereas one finds $T_d \propto N^{1/3}/\ln N$.  

6) As explicitly shown in Fig. 1, the high energy tail of the velocity distribution is not of 
Maxwell-Boltzmann type and it cannot be straightforwardly used to determine the experiment temperature 
\cite{Ensher}. 
\hspace{1cm}

\begin{center}
\epsfig{file=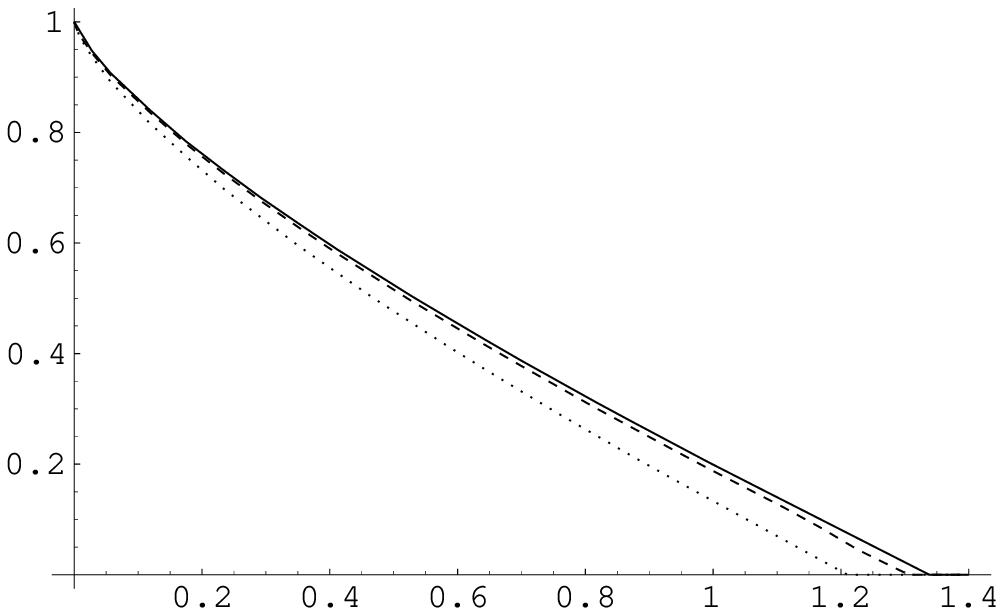,height=5cm}
\end{center}

{\noindent \footnotesize{FIG. 3. The fraction of condensate $n_0/N$ {\sl vs.} $N^{-1/3}\,
\bigl ( E/Nh\nu -3/2\bigr )$ for $N=10^4$ ($\cdot\cdot\cdot $), $10^6$ (- - -), $10^8$ (---).}} 

\hspace{1cm}
 
7) Thermo-dynamical temperature, despite its conceptual relevance, does not appear to be the 
physical parameter that best describes the process of condensation: it is difficult to measure 
in the typical experimental set-up, and as one does not have equipartition its intuitive 
meaning of $''$average energy per particle$\, ''$ is lost.  We prefer then to describe the onset  
of condensation by exhibiting the behavior of $n_0/N$ {\sl vs.} the true (dimensionless) $''$energy 
per particle$\, ''$ $E/Nh\nu$ (Fig. 3, where in abscissa we actually reported $N^{-1/3}\, 
\bigl ( E/Nh\nu -3/2\bigr )$ to point out the dependence on $N^{1/3}$), much easier to compare 
with experiments.  Our results are consistent with the experimental findings in ref. \cite{Ensher}. 
Fig. 4, of which Fig. 2 shows the derivative, provides $E/N$ as a function of $T$. This curve can be 
inverted giving $T=T(E/N)$, a relation which replaces equipartition (where it would be linear) and 
allows us to evaluate temperature dependent quantities as functions of the experimentally more accessible 
internal energy. 
\hspace{1cm}

\begin{center}
\epsfig{file=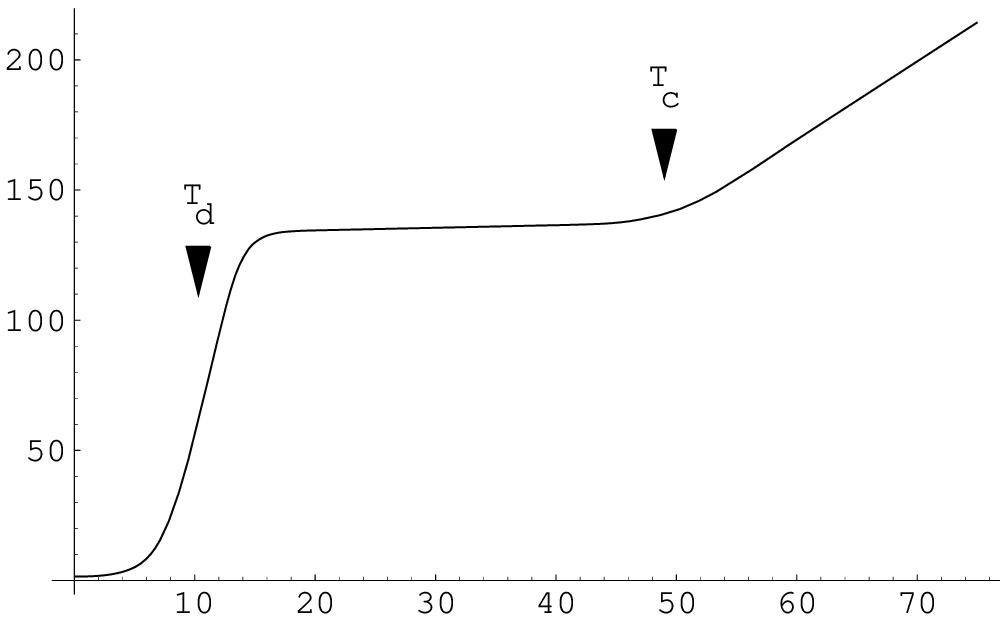,height=5cm}
\end{center}

{\noindent \footnotesize{FIG. 4. Specific internal energy $E/N$ {\sl vs.} $T$, for $N=10^6$ (units as 
in Fig. 2).}} 

\hspace{1cm}

We shall now discuss in some mathematical detail how the above features come into 
play.  The traditional textbook way (see {\sl e.g.} \cite{Hua}) of deriving the boson 
gas equilibrium distribution is the following.  Starting from Gibbs' assumption, 
in order to derive the combinatorics many single particle states (levels) of energy 
practically equal are collected together in cells (an assumption justified {\sl 
a posteriori} by the thermodynamic limit) and the particle distribution is 
straightforwardly found to be, up to normalization, given by 
\begin{eqnarray}
W\{n_i\} = \prod_i \frac{\Gamma(n_i + g)}{\Gamma(n_i+1) \Gamma(g)} \; , 
\nonumber 
\end{eqnarray}
where $g$ is the multiplicity, that with no loss of generality we assume equal for all 
cells. The equilibrium distribution, identified with that of maximum probability, is then 
obtained by finding the set $\{ n_i \}$ that maximizes (the logarithm of) $W\{n_i\}$, 
constrained by the two conditions $\sum_i n_i = N$ and $\sum_i \epsilon_i n_i = E$. One 
obtains the exact equation 
\begin{eqnarray}
\psi(n_i+g)-\psi(n_i+1) = \alpha + \beta \epsilon_i \equiv h_i \; ,  
\label{psi} 
\end{eqnarray}
where $\alpha =\alpha (g)$, dimensionless, and $\beta =\beta (g)$, with the physical dimension 
of an inverse energy, are the two Lagrange multipliers, clearly depending on the value of $g$, 
introduced in order to implement the constraints, while $\psi$ is the digamma function \cite{AbSt}. 

For large argument $\psi$ coincides with the logarithm, thus for $n_i \gg 1$ and $g\gg 1$ it 
is easy to obtain (identifying $n(\epsilon_i )$ with the finite density $n_i/g$)
\begin{eqnarray}
  n(\epsilon_i) = \frac{1}{e^{h_i}-1} \; . \label{B-E} 
\label{cont} 
\end{eqnarray} 
The latter is the Bose-Einstein distribution we are accustomed to, so well established that 
it is sometimes considered as an alternative definition of bosons. The essential assumption 
of this derivation is of course that the energy spectrum is continuous, hypothesis that does 
not hold in the presence of an harmonic potential.  

Eq. (\ref{psi}) has more general validity in that it holds true for any value of $g$ and hence also 
for the discrete case. For example, in the case of photons in a laser beam, where degeneracy of states 
is two because of helicity, it gives \cite{AbSt} 
\begin{eqnarray}
   n_i = \frac{1}{h_i} - 1 \;\, {\rm for} \;\, h_i < 1 \quad {\rm and} \quad 
   n_i = 0 \;\, {\rm for} \;\, h_i > 1 \; . 
\label{photons}  
\end{eqnarray} 

For a physical (3-dimensional) quantum harmonic oscillator, $g$ should be set equal to 1.  Indeed, even 
in the $''$isotropic$\, ''$ case, that we shall consider as an example, the experimental frequencies are 
such that $\nu_i - \nu_j \ll \nu_i$, but 
$\nu_i$ is never exactly equal to $\nu_j$; we have then a computing degeneracy only, because statistics 
remains undegenerate. However, setting $g=1$ leads for (\ref{psi}) 
to a perfectly acceptable but void equation (left hand side identically zero, implying $\alpha (1) = \beta 
(1) =0$). On the other hand: i) eq. (\ref{psi}) is defined for $g$ in the whole complex plane except for 
negative integers, and ii) in ref. \cite{CR} an alternative definition of bosons has been 
proposed in terms of coherent states of the discrete series of $\widetilde{SU}(1,1)$, where 
the highest weight of the representation is $g/2$, with $g$ any strictly positive real number. 
In this latter context $g$ is not restricted to the integer values imposed by the combinatorial 
interpretation and we can define by analytical continuation for $g \to 1$ the equation for  
undegenerate levels that we need.  $n_i \equiv n(\epsilon_i)$ results thus to be the solution to 
the equation 
\begin{eqnarray}
\psi'(n_i+1) = \alpha' + \beta' \epsilon_i \equiv h_i' \; , 
\label{dis} 
\end{eqnarray} 
where $\displaystyle{\alpha' \equiv \lim_{g\to 1} \alpha (g)/(g-1)}$ and $\displaystyle{\beta' 
\equiv \lim_{g\to 1} \beta (g)/(g-1)}$ are the new Lagrange multipliers to be determined by the 
constraints of fixed $N$ and $E$, and $\psi'$ is the trigamma function \cite{AbSt}, derivative 
of $\psi$. Eq.(\ref{dis}) is one-to-one and can be inverted, giving us the correct formula for 
discrete spectrum, to be used instead of eq.(\ref{cont}):
\begin{eqnarray}
n_i &=& [\psi']^{-1}(h_i') -1 \quad {\rm for} \quad h_i' <  \pi^2/6 
\nonumber \\ {\rm and} \qquad\quad\quad &~& \label{inv} \\  n_i &=& 0 \quad {\rm for} \quad h_i'  
\geq \pi^2/6 \; ,   
\nonumber 
\end{eqnarray}
$[\psi']^{-1}$ denoting the inverse function of $\psi'$. $\pi^2/6 \equiv \psi' (1)$. For $h' \to 0$ eq. 
(\ref{inv}) has the same asymptotic behavior as (\ref{B-E}) : $n_i \to {h'}^{-1} - \frac{1}{2}$.  The two 
formulas (\ref{cont}) and (\ref{inv}) exhibit a similar structure, characteristic of all values of 
$g$, with a pole (with residue 1) at argument zero and a zero for large values of the argument (compare  
eq. (\ref{photons})). The discrete description, however, does not require an {\sl ad hoc} delta 
function at the ground state, which is instead discussed on the same footing as all the others. 
In fact one may assign first the values of $N$ and $E$; from these the two Lagrange multipliers $\alpha' 
= \alpha' (N,E)$, $\beta' = \beta' (N,E)$ are then obtained from the constraints using for $n_i$ the 
defining equation (\ref{inv}); then, inserting them in (\ref{inv}) 
itself, one gets $n_i \equiv n_i (\alpha' ,\beta' ) = n_i(N,E)$ (see Figs. 1 and 3). 

As condensation derives from the existence of the pole combined with the feature that multiplicity 
increases quadratically for the excited states, the continuum formula (\ref{cont}) is correct in 
predicting the collapse of the atoms in the fundamental level, but it is unable to describe what 
happens for $T < T_c$.  The subtlety here is that the very concept of temperature, as defined in the 
theory of gases, cannot be straightforwardly extended to the discrete spectrum case.

Numerical calculations performed for a physical 3-dimensional $''$isotropic$\, ''$ 
quantum harmonic oscillator with $N$ ranging from $10^2$ to $10^{12}$, made for simplicity 
approximating $[\psi']^{-1}$ by fixing only pole and zero, 
\begin{eqnarray}
 n_i = \frac{1}{h_i} - \frac{6}{\pi^2} \;\, {\rm for} \;\, h_i < \pi^2/6 \; {\rm and} \; 
   n_i = 0 \;\, {\rm for} \;\, h_i > \pi^2/6 \; ,   
\nonumber 
\end{eqnarray} 
show indeed that, for $E/N$ large, $\pi^{2}/6 - \alpha' \propto {\displaystyle{\left ( E/N \right 
)}}^{-3}$ while $\beta' \propto {\displaystyle{\left ( E/N \right )}}^{-4}$. The internal energy 
is therefore not proportional to the inverse Lagrange multiplier ${\beta'}^{-1}$ but to the more complex expression 
$( \pi^2/6 - \alpha' )\, {\beta'}^{-1}$.  Temperature can then no longer be obtained, as in the continuum, 
by $T = \bigl ( k_B \beta' \bigr )^{-1}$ and equipartition of energy must be demonstrated. The only way to 
proceed is to adopt the basic definition of temperature based on entropy.  In order to do so, we use 
Shannon's definition of entropy in terms of probabilities 
\begin{eqnarray}
S = - k_B \, \sum \, p_i \, \ln p_i 
\nonumber 
\end{eqnarray}
where we set $p_i \equiv n_i/N$, recalling that the $n_i(N,E)$'s are now known.  At fixed $N$ 
we thus straightforwardly obtain $S = S(E)$.  Simple numerical operations finally give the temperature
\begin{eqnarray}
T \equiv T(E) = \left [\frac{\partial S}{\partial E}\right ]_N^{-1} \; . 
\nonumber 
\end{eqnarray}    
Once more the function is one-to-one, and from $T = T(E)$ we can obtain $E = E(T)$ (see Fig. 4), whence 
the specific heat $\displaystyle{C \equiv \frac{1}{N}\, \left [ \frac{\partial E}{\partial T} \right ]_N}$ 
is obtained by (numerical) differentiation (see Fig. 2).  Also $\mu$, which is no longer  
related in a simple way to $\alpha'$, can be calculated in analogous way, $\displaystyle{\mu \equiv 
- \left [ \frac{\partial E}{\partial N} \right ]_S}$, and all the results presented above are thus 
obtained.    

It should be pointed out that the theory presented leads not only to a set of specific 
behaviors, but it determines as well the measurable quantities: specific heat, filling 
order parameter, chemical potential, etc., all of which can be experimentally measured 
and are related to the characteristics of the experiment ($N$, $E$, $\nu$). To this latter 
effect, it should be pointed out that in the experiments the harmonic potential is usually 
anisotropic: extension of our results to the corresponding physical parameters is 
straightforward. Whether or not the features predicted theoretically are correct or 
not can therefore be tested by direct measurements of sufficiently high sensitivity. This 
feedback from the laboratory is crucial before further refinements of our analysis, such as 
{\sl e.g.} inclusion of many body effects, is carried over.

\end{multicols}
\end{document}